\RequirePackage{fix-cm}
\documentclass[twocolumn,epjc3]{svjour3}  
\smartqed  
\RequirePackage{graphicx}
\RequirePackage{latexsym}
\usepackage{amsmath,amssymb}
\usepackage{lineno}
\usepackage{bbding}
\journalname{Eur. Phys. J. C}
\begin{document}

\title{The $v^{1/3}_{3}/v^{1/2}_{2}$ ratio in PbAu collisions at $\sqrt{s_{\mathrm{NN}}} = $ 17.3~GeV - a hint of a hydrodynamic behavior
}
\titlerunning{The $v^{1/3}_{3}/v^{1/2}_{2}$ ratio ...}        

\author{
  CERES/NA45 Collaboration \\
  D.~Adamov{\'a}\thanksref{addr1}
  \and
  G.~Agakishiev\thanksref{addr2}
  \and
  A.~Andronic\thanksref{addr3}
  \and
  D.~Anto{\'n}czyk\thanksref{addr4}
  \and
  H.~Appelsh{\"a}user\thanksref{addr4}
  \and
  V.~Belaga$^{\dag}$\thanksref{addr2}
  \and
  J.~Biel\v{c}\'{\i}kov{\'a}\thanksref{addr5,addr6,addr7}
  \and
  P.~Braun-Munzinger\thanksref{addr8}
  \and
  O.~Busch$^{\dag}$\thanksref{addr9}
  \and
  A.~Cherlin\thanksref{addr10}
  \and
  S.~Damjanovi{\'c}\thanksref{addr9}
  \and
  T.~Dietel\thanksref{addr3}
  \and
  L.~Dietrich\thanksref{addr9}
  \and
  A.~Drees\thanksref{addr11}
  \and
  W.~Dubitzky\thanksref{addr9}
  \and
  S.~I.~Esumi\thanksref{addr9,addr12}
  \and
  K.~Filimonov\thanksref{addr9,addr13}
  \and
  K.~Fomenko\thanksref{addr2}
  \and
  Z.~Fraenkel$^{\dag}$\thanksref{addr10}
  \and
  C.~Garabatos\thanksref{addr8}
  \and
  P.~Gl{\"a}ssel\thanksref{addr9}
  \and
  G.~Hering\thanksref{addr8}
  \and
  J.~Holeczek\thanksref{addr8}
  \and
  M.~Kalisky\thanksref{addr3}
  \and
  G.~Krobath\thanksref{addr9}
  \and
  V.~Kushpil\thanksref{addr1}
  \and
  A.~Maas\thanksref{addr8}
  \and
  A.~Mar\'{\i}n\thanksref{addr8}
  \and
  J.~Milo\v{s}evi{\'c}\thanksref{e1,addr9,addr15,addr16}
  \and
  D.~Mi{\'s}kowiec\thanksref{addr8}
  \and
  Y.~Panebrattsev\thanksref{addr2}
  \and
  Z.~Paul\'{i}nyov\'{a}\thanksref{addr17}
  \and
  O.~Petchenova\thanksref{addr2}
  \and
  V.~Petr{\'a}\v{c}ek\thanksref{addr9,addr7}
  \and
  S.~Radomski\thanksref{addr9}
  \and
  J.~Rak\thanksref{addr5}
  \and
  I.~Ravinovich\thanksref{addr10}
  \and
  P.~Rehak$^{\dag}$\thanksref{addr18}
  \and
  H.~Sako\thanksref{addr8}
  \and
  W.~Schmitz\thanksref{addr9}
  \and
  S.~Schuchmann\thanksref{addr4}
  \and
  S.~Sedykh\thanksref{addr8}
  \and
  S.~Shimansky\thanksref{addr2}
  \and
  J.~Stachel\thanksref{addr9}
  \and
  M.~\v{S}umbera\thanksref{addr1}
  \and
  H.~Tilsner\thanksref{addr9}
  \and
  I.~Tserruya\thanksref{addr10}
  \and
  G.~Tsiledakis\thanksref{addr8}
  \and
  J.\thinspace P.~Wessels\thanksref{addr3}
  \and
  T.~Wienold\thanksref{addr9}
  \and
  J.\thinspace P.~Wurm$^{\dag}$\thanksref{addr5}
  \and
  S.~Yurevich\thanksref{addr9,addr2}
  \and
  V.~Yurevich\thanksref{addr2}
}

\thankstext{e1}{e-mail: Jovan.Milosevic@cern.ch}


\institute{
  Nuclear Physics Institute, Czech Academy of Sciences, 25068 \v{R}e\v{z}, Czech Republic \label{addr1}
  \and
  Joint Institute of Nuclear Research, Dubna, 141980 Moscow Region, Russia \label{addr2}
  \and
  Institut f\"ur Kernphysik, Universit{\"a}t M{\"u}nster, 48149 M\"unster, Germany \label{addr3}
  \and
  Institut f\"ur Kernphysik, Johann Wolfgang Goethe-Universit{\"a}t Frankfurt, 60438 Frankfurt, Germany \label{addr4}
  \and
  Max-Planck-Institut f{\"u}r Kernphysik, 69117 Heidelberg, Germany \label{addr5}
  \and
  \emph{Present Address:} Nuclear Physics Institute, Czech Academy of Sciences, 25068 \v{R}e\v{z}, Czech Republic \label{addr6}
  \and
  \emph{Present Address:} Czech Technical University in Prague, Faculty of Nuclear Sciences and Physical Engineering, 11519 Prague, Czech Republic \label{addr7}
  \and
  Research Division and Extreme Matter Institute (EMMI), GSI Helmholtzzentrum f\"ur Schwerionenforschung, 64291 Darmstadt, Germany \label{addr8}
  \and
  Physikalisches Institut, Universit{\"a}t Heidelberg, 69120 Heidelberg, Germany \label{addr9}
  \and
  Department of Particle Physics, Weizmann Institute, Rehovot, 76100 Israel \label{addr10}
  \and
  Department for Physics and Astronomy, SUNY Stony Brook, NY 11974, USA \label{addr11}
  \and
  \emph{Present Address:} Institute of Physics, University of Tsukuba, Tsukuba, Japan \label{addr12}
  \and
  \emph{Present Address:} Physics Department, University of California, Berkeley, CA 94720-7300, USA \label{addr13}
  \and
  \emph{Present Address:} ``VIN\v{C}A'' Institute of Nuclear Science - National Institute of the Republic of Serbia, University of Belgrade, Mike Petrovi\'{c}a Alasa 12-14, Vin\v{c}a 11351, Belgrade, Serbia \label{addr15}
  \and
  \emph{Also at:} Strong-coupling Physics International Research Laboratory, Huzhou University, China \label{addr16}
  \and
  Faculty of Science, P. J. \v{S}af\'{a}rik University, Ko\v{s}ice, Slovakia \label{addr17}
  \and
  Instrumentation Division, Brookhaven National Laboratory, Upton, NY 11973-5000, USA \label{addr18}
}

\date{Received: date / Accepted: date}

\maketitle

\begin{abstract}
The Fourier harmonics, $v_2$ and $v_3$ of negative pions are measured at center-of-mass energy per nucleon pair of $\sqrt{s_{\mathrm{NN}}}$= 17.3~GeV around midrapidity by the CERES/NA45 experiment at the CERN SPS in 0--30\% central PbAu collisions with a mean centrality of 5.5\%. The analysis is performed in two centrality bins as a function of the transverse momentum $\mathrm{p_{\mathrm{T}}}$ from 0.05~GeV/$c$ to more than 2~GeV/$c$. This is the first measurement of the $v^{1/3}_{3}/v^{1/2}_{2}$ ratio as a function of transverse momentum at SPS energies, that reveals, independently of the hydrodynamic models, hydrodynamic behavior of the formed system. For $\mathrm{p_{\mathrm{T}}}$ above 0.5~GeV/$c$, the ratio is nearly flat in accordance with the hydrodynamic prediction and as previously observed by the ATLAS and ALICE experiments at the much higher LHC energies. The results are also compared with the SMASH-vHLLE hybrid model predictions, as well as with the SMASH model applied alone.
\end{abstract}

\section{Introduction}
\label{introduction}
At sufficiently high energy density as achieved in ultra-relativistic heavy-ion collisions, a dense and hot system of strongly coupled quarks and gluons called Quark-Gluon-Plasma (QGP) is created~\cite{Heinz:2000bk,BRAHMS:2004adc,PHOBOS:2004zne,STAR:2005gfr,PHENIX:2004vcz,Busza:2018rrf,Braun-Munzinger:2015hba}. Anisotropic pressure gradients, built early in the collision, cause a collective azimuthally anisotropic expansion of the QGP that converts initial spatial anisotropy into a momentum anisotropy that was observed first at the BNL AGS (E877)~\cite{E877:1994plr}, then at the CERN SPS (NA49, WA98, CERES)~\cite{NA49:1996ytw,WA98:1998ucl,CERES:1998nee} and finally at Relativistic Heavy Ion Collider (RHIC) and Large Hadron Collider (LHC) experiments~\cite{ALICE:2010suc,ALICE:2011ab,ALICE:2014wao,ALICE:2016ccg,ATLAS:2011ah,ATLAS:2012at,ATLAS:2013xzf,CMS:2012xss,CMS:2012zex,CMS:2013wjq,CMS:2013bza}. This azimuthal anisotropy of particles emitted in the final state can be used to explore the hydrodynamic behavior of the hot and dense systems created in such collisions. The almond-like shape of the overlapping region in a non-central nucleus-nucleus collision causes the appearance of the elliptic flow ($v_{2}$) anisotropy~\cite{Ollitrault:1992bk}. Moreover, the positions of nucleons in the colliding nuclei fluctuate. These event-by-event fluctuations have a significant influence on the QGP expansion~\cite{Alver:2010gr} and cause the appearance of higher-order anisotropies. The most pronounced is the triangular flow ($v_{3}$) that is measured at both RHIC~\cite{PHOBOS:2006dbo,PHOBOS:2007vdf,PHENIX:2011yyh,Sun:2014yqi,STAR:2016vqt} and the LHC~\cite{ALICE:2011ab,ATLAS:2012at,CMS:2013wjq,CMS:2013bza} in central and non-central collisions.

At the high collision energies achieved at RHIC and LHC, large $v_2$ values are observed that agree well with predictions of relativistic hydrodynamics~\cite{Huovinen:2006jp} without dissipation, suggesting that elliptic flow is built early in the evolution of the QGP, and that the QGP behaves as a nearly perfect liquid, see the reviews in~\cite{Gyulassy:2004zy,Hirano:2005xf}. The measured $v_2$ is also well described with a hybrid calculation that treats the QGP by ideal hydrodynamics~\cite{Ollitrault:1992bk} and the late stages of the collision by a hadronic cascade model~\cite{Hirano:2010jg}. The description can be refind by taking viscosity into account~\cite{Teaney:2001gc,Romatschke:2007mq}. The viscosity makes the $v_2$ magnitudes somewhat smaller with respect to the ideal hydrodynamic limit. The viscous hydrodynamic models like MUSIC~\cite{Schenke:2010nt,Schenke:2010rr} and VISHNU~\cite{Shen:2014vra} have been developed to describe azimuthal anisotropies. A very good agreement between the IP Glasma+MUSIC~\cite{Gale:2012rq}, as well as the VISHNU model predictions~\cite{Song:2011hk}, and the experimentally measured azimuthal anisotropies at RHIC and LHC are obtained. Relativistic hydrodynamic models achieved to give semiquantitative constraints on specific viscosity ($\eta/s$), but with large uncertainties. Comprehensive model-data analyses that use a Bayesian approach allow a temperature-dependent specific viscosity~\cite{Bernhard:2019bmu}. The obtained values of $\eta/s$, depending on the temperature, vary from 0.1 up to 0.2. Such small values show that QGP is a strongly-coupled nearly perfect fluid~\cite{Gyulassy:2004zy,Shuryak:2004kh}. Contrary to the elliptic flow, the triangular flow is nearly independent of centrality, and can be described using viscous hydrodynamics and transport models too. Triangular flow is a sensitive probe of initial geometry fluctuations and viscosity~\cite{Alver:2010dn}.

The $v_2$ magnitudes measured at the Super Proton Synchrotron (SPS) energy, $\sqrt{s_{\mathrm{NN}}}=17.3$~GeV, are significantly lower than those measured at the top RHIC and LHC energies. Except for the most central collisions~\cite{CERES:2012hbp}, the transverse momentum flow data $v_2(p_{\rm T})$ at SPS~\cite{NA49:2003njx,CERESNA45:2003pat,WA98:2004dfx}, although very similar in shape to the RHIC and LHC data, are below predictions of ideal hydrodynamics~\cite{Huovinen:2001cy}. The inability of the ideal hydrodynamics to describe the elliptic flow at the top SPS energy has been ascribed to insufficient number densities at very early collision stages~\cite{Heinz:2004et} and strong dissipative effects at the late hadronic stages~\cite{Gyulassy:2004zy,Hirano:2005xf,Teaney:2003kp,Niemi:2011ix}. The flow measurements have been also performed within the Beam Energy Scan (BES) program at the RHIC from the top incident energy of $\sqrt{s_{\mathrm{NN}}}=200$~GeV down up to the $\sqrt{s_{\mathrm{NN}}}=7.7$~GeV~\cite{STAR:2016ydv,STAR:2015rxv}. Also, at the RHIC have been measured various correlations between the Fourier $v_{n}$ harmonics, among them the ratio $v_{4}/v^{2}_{2}$ as a function of $p_{\rm T}$ and centrality~\cite{STAR:2004jwm,PHENIX:2010tme}.

In this paper, we present the first result of negative pions $v^{1/3}_{3}/v^{1/2}_{2}$ ratio as a function of transverse momentum in PbAu collisions at the SPS at a nucleon-nucleon center-of-mass energy of $\sqrt{s_{\mathrm{NN}}} = $ 17.3~GeV. The ana\-lysis is based on previously published $v_{2}$ and $v_{3}$ data~\cite{CERES:2012hbp,CERESNA45:2016rra}, where the $v_{2}$ data have been rebinned. This reanalysis of the data provides a novel way of highlighting similarities with the flow phenomena seen at RHIC and LHC. The obtained results reveal, independently of the hydrodynamic models, a new interpretation of the SPS data which indicates hydrodynamic behavior of the system formed in PbAu collisions already at top SPS energy. The results are compared with the ATLAS and the ALICE measurements and show a very similar behavior. The results are also compared with predictions from a hybrid model~\cite{Karpenko:2015xea} consisting of SMASH (Simulating Many Accelerated Strongly-interacting Handrons)~\cite{SMASH:2016zqf}, a hadron transport approach suitable for modelling heavy-ion collisions, combined with the vHLLE 3+1D viscous hydrosolver~\cite{Karpenko:2013wva} that simulates the QGP expansion, as well as with predictions from the SMASH model applied alone. The paper is organized as follows. Some details concerning the experiment and the data sample are given in section~\ref{experiment}. The data analysis and the obtained results are presented in section~\ref{results}. A short summary is given in section~\ref{summary}.

\section{Experiment and data sample}
\label{experiment}

In the year 2000, a sample of $30\cdot 10^{6}$ mainly central PbAu events was collected with the upgraded CERES/ NA45 spectrometer at the top SPS center-of-mass energy per nucleon pair of $\sqrt{s_{\mathrm{NN}}} = $ 17.3~GeV. The spectrometer covers the polar angle acceptance of $7.7^\circ \leq \vartheta \leq 14.7^\circ$, corresponding to a pseudorapidity range $2.05 \leq \eta \leq 2.70$ located near midrapidity ($y_{\rm mid}=2.91$). The CERES/NA45 spectrometer has axial symmetry with respect to the beam direction. As it covers the full $2\pi$ azimuthal acceptance, it is very well suited for measurements of the Fourier harmonics of azimuthal distributions. A detailed description of the CERES/NA45 experiment is given in \cite{Marin:2004fx}.

Momentum determination is provided by the radial-drift Time Projection Chamber (TPC)~\cite{CERES:2008asn} that is operated inside an axially symmetric magnetic field with a maximum radial component of 0.5~T. Negative pions are identified by the differential energy loss d$E$/d$x$ in the TPC. For vertex reconstruction and tracking outside the magnetic field, two radial Silicon Drift Detectors (SDD)~\cite{Holl:1996fp} are placed at 10 and 13~cm downstream of a segmented Au target. Negative pions are reconstructed by matching track segments in the SDD doublet and in the TPC. Depending on pion momentum, the relative momentum resolution varies between 2\% and 8\% for pion momenta of 0.05 to 2.0 GeV/c.

\begin{figure}[h!]
\centering
\includegraphics[width=8.5cm] {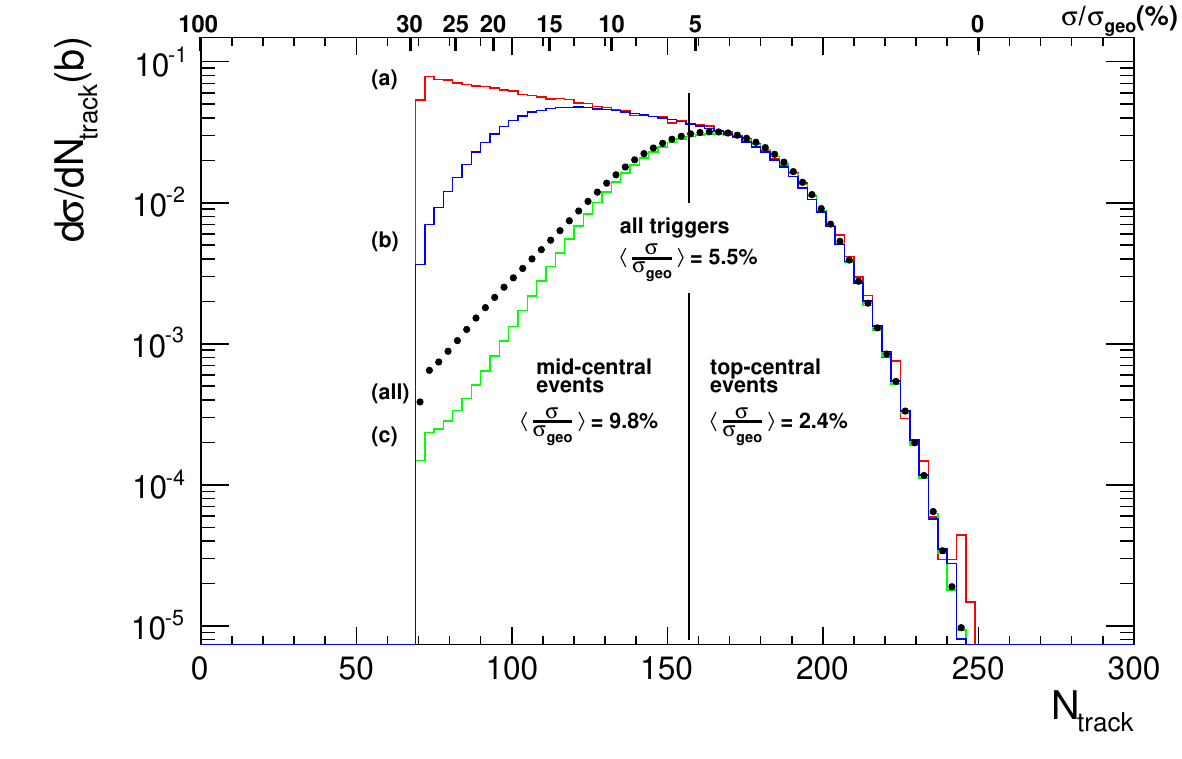}
  \caption{TPC track density for the trigger mix within (0 -- 30\%) centrality. The mix consists of three components: (a) minimum-bias (0.5\%), (b) semicentral (8.3\%), and (c) central (91.2\%), where the parentheses represent the percentage fractions in the mix. The mix of all triggers, with a resulting mean centrality of 5.5~\%, is labeled as `all triggers' and displayed by closed black circles. The vertical axis gives the differential cross-section expressed in barns (b). The $\langle\sigma/\sigma_{geo}\rangle$ axis on top gives the fraction of the total inelastic cross section. The figure is taken from Ref.~\cite{CERESNA45:2016rra}.
 \label{fig:mult_NPA}}
 \end{figure}

A mix of three triggers designed to enhance central events was used for data collection in the range \mbox{0 -- 30\%} of $\sigma/\sigma_{\rm geo}$ with an average centrality of 5.5\% in the data sample. The multiplicity distribution for all triggers combined (`all triggers') is shown in Fig.~\ref{fig:mult_NPA} by the closed black symbols. At low multiplicities, it strongly deviates from the minimum-bias distribution labeled with $(a)$. Beside minimum-bias data, which contribute with only 0.5\% of all events, a semi-central trigger, labeled with $(b)$, contributes 8.3~\% to the total. The biggest share of data, 91.2\%, is collected with a most central trigger labeled with $(c)$ in Fig.~\ref{fig:mult_NPA}. The results presented in this paper correspond to `top-central' and `mid-central' triggers by selecting $N_{track} > 159$ and $N_{track} \leq 159$, with weighted mean centralities of 2.4\% and 9.8\%, respectively.

\section{Analysis and results}
\label{results}

The performed analysis uses the event-plane method \cite{Voloshin:1994mz,Poskanzer:1998yz}. The event plane is defined experimentally at high energies by the beam direction and the direction of maximum outgoing particle density, and it is characterized by the event-plane angle $\Psi_{n}$. Each Fourier harmonic order $n$ has its own event plane. The angle $\Psi_{n}$ is calculated as:
\begin{equation}
  \label{eq:Phin}
  \begin{aligned}
\Psi_{n}=\frac{1}{n}\arctan\frac
{\displaystyle\sum\limits_{i=0}^Mw_{i}(p_{\mathrm{T}i})
  \sin(n\phi_i)}{\displaystyle\sum\limits_{i=0}^Mw_{i}(p_{\mathrm{T}i})\cos(n\phi_i)}, \; n=2,3,
\end{aligned}
\end {equation} 
where $\phi_i$ is the azimuthal angle of the $i^{th}$ particle out of $M$ particles, with transverse momentum $p_{\mathrm{T}i}$, used for event-plane reconstruction, and $w_{i}(p_{\mathrm{T}i})$ are weights used to optimize the event-plane resolution. In order to avoid a trivial autocorrelation effect, and some contribution from short-range correlations, the $\phi$ coordinate is divided into 100 adjacent equal slices spanning the whole $\phi$ range. A set made from the each fourth slice forms a 'sliced subevent'. So, the whole event is divided into 4 sliced subevents each consisting of 25 slices. These four subevents are denoted as $a$, $b$, $c$ and $d$. The $n^{th}$ order event plane is reconstructed by employing tracks that belong to a given sliced subevent. In order to correct for local detector inefficiency, a shifting and flattening procedure is applied (for more details see~\cite{Jovan06}) to ensure an azimuthally isotropic event-plane distribution. The anistropy is measured by correlating particles from a given sliced subevent with the event plane obtained from the non-adjacent sliced subevent, i.e $a - c$ and $b - d$ combinations. The $v_{n}$ coefficients are then measured with respect to the $n^{th}$ order event plane. The raw $v_{n}$ coefficients are corrected for the finite event-plane resolution.
  
The analysis is performed for negative pions\footnote{Using the d$E$/d$x$, positive pions could not be well enough separated from protons.}. As identical bosons, they are correlated due to the Hanbury Brown \& Twiss (HBT) effect~\cite{HanburyBrown:1954amm}. This correlation results in a spurious flow~\cite{Dinh:1999mn}. A standard procedure, that uses the Bertsch-Pratt parametrization described in~\cite{Dinh:1999mn}, is applied to suppress the non-flow contribution coming from the HBT effect. Details about the procedure can be found in~\cite{CERES:2012hbp,CERESNA45:2016rra,Jovan06}.

The most important source of systematic uncertainty stems from the event plane determination. To contrast to the proces analysis, we reduce and estimate this uncertainty by employing two methods. Beside the 'sliced subevent' method, the data have been analized using the subevent method too. In the subevent method, the event is divided into two subevents, one formed from negative pion, and another one from positive pion candidates selected using a band that corresponds to $\pm$1.5$\sigma$ confidence level around the nominal $dE/dx$ energy loss for negative, and for positive pions calculated by using the Bethe-Bloch formula respectively. Again, the autocorrelation effect is removed by correlating the pions from one subevent to the event plane constructed from the other one. The advantage of this approach with respect to the ’sliced subevent’ method is that the event-plane resolution is better by a factor of about $\sqrt{2}$ due to the twice larger multiplicity used for the event-plane reconstruction~\cite{Jovan06}. Then, the systematic uncertainty is calculated as an absolute difference between the $v_{n}$ ($n = $ 2, 3) magnitudes calculated from the two methods: the 'sliced subevent' method and the subevent method. At $p_{\mathrm{T}}$ smaller than 0.8~GeV/c, a significant contribution to the systematic uncertainty comes from the correction of the HBT effect too. That part of systematic uncertainty is calculated in the same way as it was done in~\cite{CERES:2012hbp} and in ~\cite{CERESNA45:2016rra}. The maximal absolute systematic uncertainty is about 0.4\% at $p_{\mathrm{T}}$ around 0.3~GeV/c, where the HBT effect is largest~\cite{CERESNA45:2016rra}. For $p_{\mathrm{T}}$ above 0.8~GeV/c, this effect is negligible. The final systematic uncertainty is obtained by summing up in quadrature the uncertainties from the difference between the two methods and the one that comes from the HBT effect. As the uncertainties are uncorrelated, based on the error propagation, the systematic uncertainties for the ratio $v^{1/3}_{3}/v^{1/2}_{2}$ are calculated from the corresponding values for the $v_{2}$ and $v_{3}$ coefficients. The relative systematic uncertainties are large for the ratio $v^{1/3}_{3}/v^{1/2}_{2}$, especially for the 'top-central' events class, due to the fact that at lowest $p_{\mathrm{T}}$ the flow magnitudes are close to zero. Corresponding values go from 13\% at $p_{\mathrm{T}}$ around 1.0~GeV/c up to about 100\% at the lowest $p_{\mathrm{T}}$ bin. Thus, the relative systematic uncertainties of the ratio $v^{1/3}_{3}/v^{1/2}_{2}$ dominate over the statistical ones in the low $p_{\mathrm{T}}$ bins and become about 10\% in higher $p_{\mathrm{T}}$ bins for the 'top-central' events class. The overall absolute uncertainties are rather tiny.

\begin{figure}[h!]
  \centering
  \includegraphics[width=9.2cm]{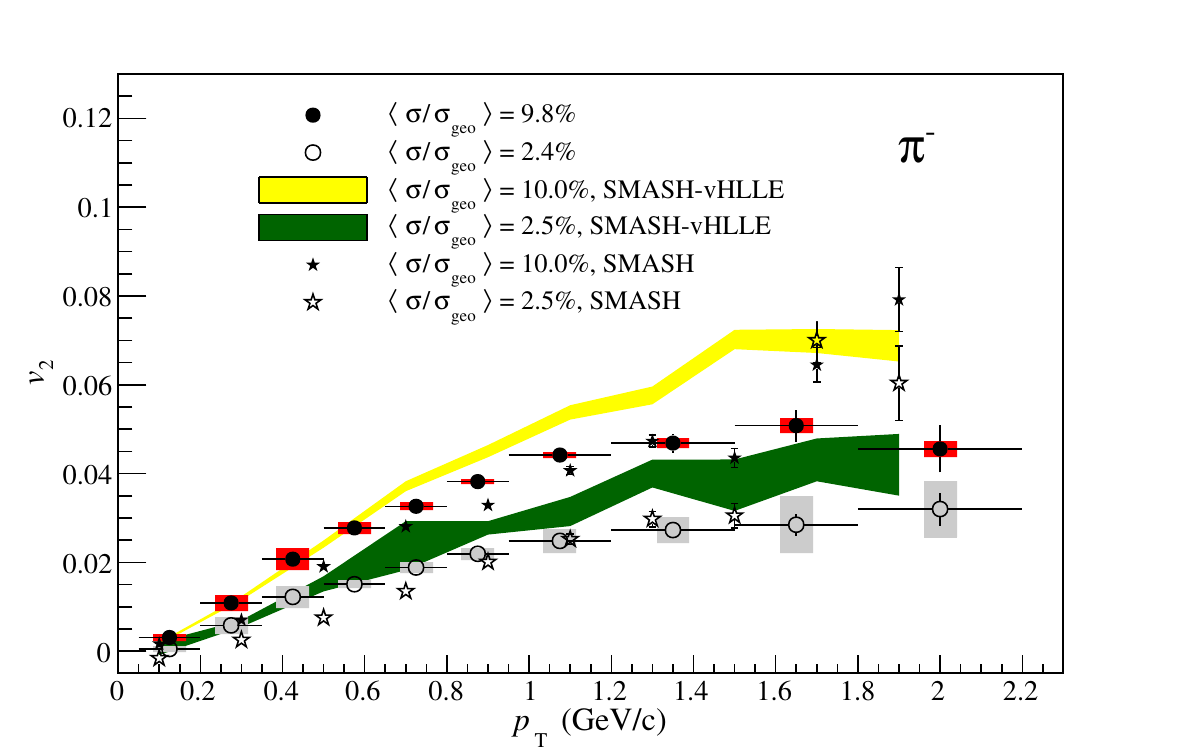}
  \caption{The transverse momentum dependence of the elliptic flow coefficient, $v_{2}$, for negative pions in PbAu collisions at $\sqrt{s_{\mathrm{NN}}} = $ 17.3~GeV in two centrality classes. The statistical uncertainties are represented with the vertical bars, while systematic ones are indicated by red and gray rectangles. Also shown are model predictions with their statistical uncertainties as a yellow and green bands (SMASH-vHLLE) and the vertical bars (SMASH).
  \label{fig:2}}
  \end{figure}

Figure~\ref{fig:2} displays the HBT-corrected $v_{2}$ values of negative pions for two centrality classes in PbAu collisions at $\sqrt{s_{\mathrm{NN}}} = $ 17.3~GeV. The $v_{2}$ values of negative pions are taken from~\cite{CERES:2012hbp} and rebinned. In the same figure results from a hybrid model~\cite{Karpenko:2015xea} combining the vHLLE viscous hydrosolver~\cite{Karpenko:2013wva} with the hadronic transport model SMASH~\cite{SMASH:2016zqf} are plotted. The $v_{n}$ coefficients are calculated using the event-plane method. The centrality ranges for the hydrodynamic calculation are comparable with the experimental ones. The model predictions are calculated for negative pions within $-1 < \eta < 1$ that is close to the experimental acceptance. The value of the specific viscosity used in the vHLLE model is 0.15. For $p_{\mathrm{T}}$ above about 0.6 GeV/c, in both centrality classes the hydrodynamic prediction overestimates the data. The SMASH model data are analyzed using the same event-plane method as in the case of the SMASH-vHLLE model, as well as the same acceptance. In difference to SMASH-vHLLE, the SMASH model alone predicts that $v_{2}$ is somewhat below the experimental data for $p_{\mathrm{T}} <$ 1 GeV/c. At high $p_{\mathrm{T}}$, due to the limited statistics in the simulations, $v_{2}$ fluctuates a lot.

\begin{figure}[h!]
  \centering
  \includegraphics[width=9.2cm]{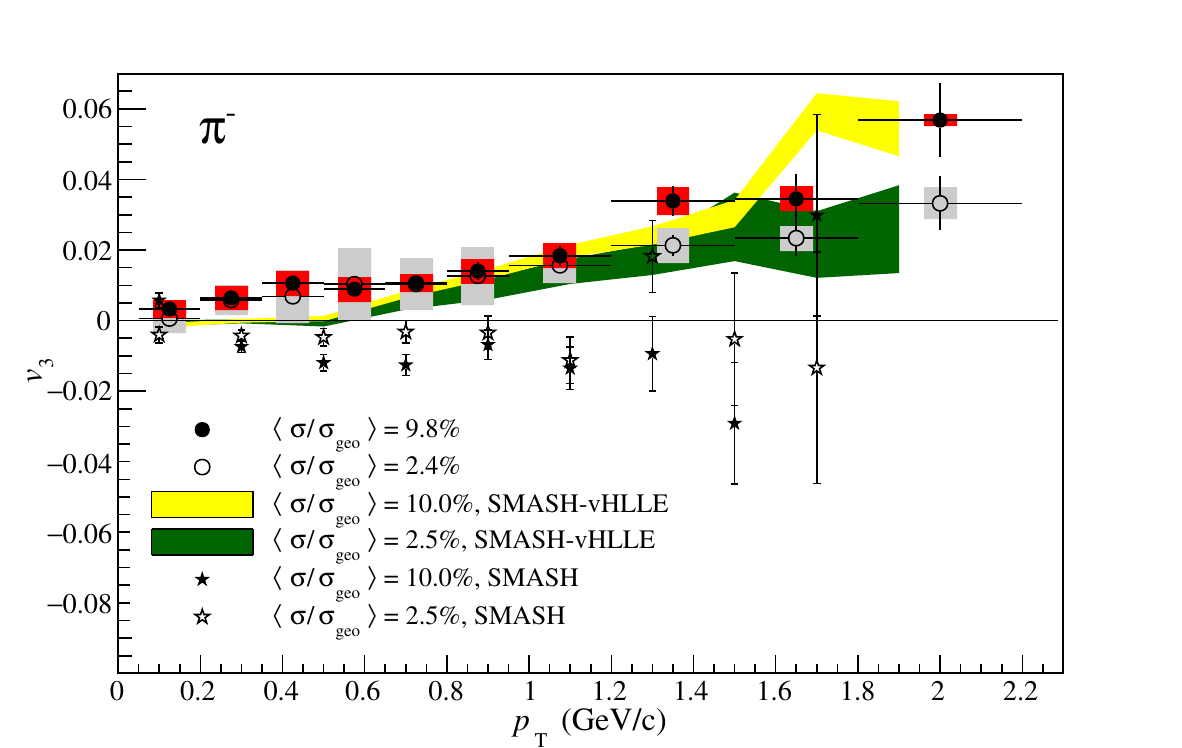}
  \caption{The transverse momentum dependence of the triangular flow coefficient, $v_{3}$, for negative pions in PbAu collisions at $\sqrt{s_{\mathrm{NN}}} = $ 17.3~GeV in two centrality classes. The statistical uncertainties are represented with the vertical bars, while systematic ones are indicated by red and gray rectangles. Also shown are model predictions with their statistical uncertainties as a yellow and green bands (SMASH-vHLLE) and the vertical bars (SMASH).
\label{fig:3}}
\end{figure}

Figure~\ref{fig:3} shows the HBT corrected $v_{3}$ values of negative pions for the same two centrality classes in PbAu collisions at $\sqrt{s_{\mathrm{NN}}} = $ 17.3~GeV. The $v_{3}$ values of negative pions are taken from~\cite{CERESNA45:2016rra}. As in Fig.~\ref{fig:2}, the results obtained from the above mentioned hybrid model are plotted together with the experimental data. In contrast to the case of the elliptic flow, the hydrodynamic calculations are in a good agreement with the data also for $p_{\mathrm{T}}$ above about 0.8 GeV/c. For $p_{\mathrm{T}} <$ 0.8 GeV/c, data tend to be above the hydrodynamic prediction for both centrality classes. In difference to the SMASH-vHLLE and the experimental results, the SMASH model alone gives small but negative $v_{3}$ that prevents calculation of the ratio $v^{1/3}_{3}/v^{1/2}_{2}$  \footnote{For clarity of the figure, the points for $v_{3}$ values for both centrality bins at $p_{\mathrm{T}} = $ 1.9~GeV/c are not included in the figure. Their values are -0.158$\pm$0.048 and -0.142$\pm$0.053 for $\langle \sigma/\sigma_{geo} \rangle$ = 9.8\% and $\langle \sigma/\sigma_{geo} \rangle$ = 2.4\% respectively.}. So, a positive $v_{3}$ is obtained only in a combination of hydrodynamics vHLLE calculations and the SMASH model (SMASH-vHLLE). A long hydrodynamic evolution of the QGP dominates the collision dynamics.

In the hydrodynamics, at $p_{\mathrm{T}} \lesssim$ 2 GeV/c, the $v_{n}(p_{\mathrm{T}})$ behave as power-law functions of $p_{\mathrm{T}}$~\cite{Alver:2010dn,Borghini:2005kd}. This can be expressed as
\begin{equation}
  \label{eq:Pow-low}
  \begin{aligned}
    v_{n}(p_{\mathrm{T}})=c_{n}p_{\mathrm{T}}^{n/m}, \; n=2,3,...,
  \end{aligned}
\end {equation}
where $c_{n}$ is a coefficient of proportionality that depends on the order $n$, and m is a fixed number\footnote{The coefficient $c_{n}$ and number $m$ can be found from the corresponding fit to the $v_{n}(p_{\mathrm{T}})$ distribution.} independent of $n$. In this case, the ratio $v^{1/3}_{3}/v^{1/2}_{2}$ becomes a $p_{\mathrm{T}}$ independent number $c^{1/3}_{3}/c^{1/2}_{2}$. The ratio depends on centrality. Similar scaling ratios $v^{1/n}_{n}/v^{1/2}_{2}$, $n > 2$, as a function of centrality, have been already measured at RHIC~\cite{PHENIX:2010tme} and the LHC~\cite{ATLAS:2012at}.

\begin{figure}[h]
  \centering
\includegraphics[width=9.2cm]{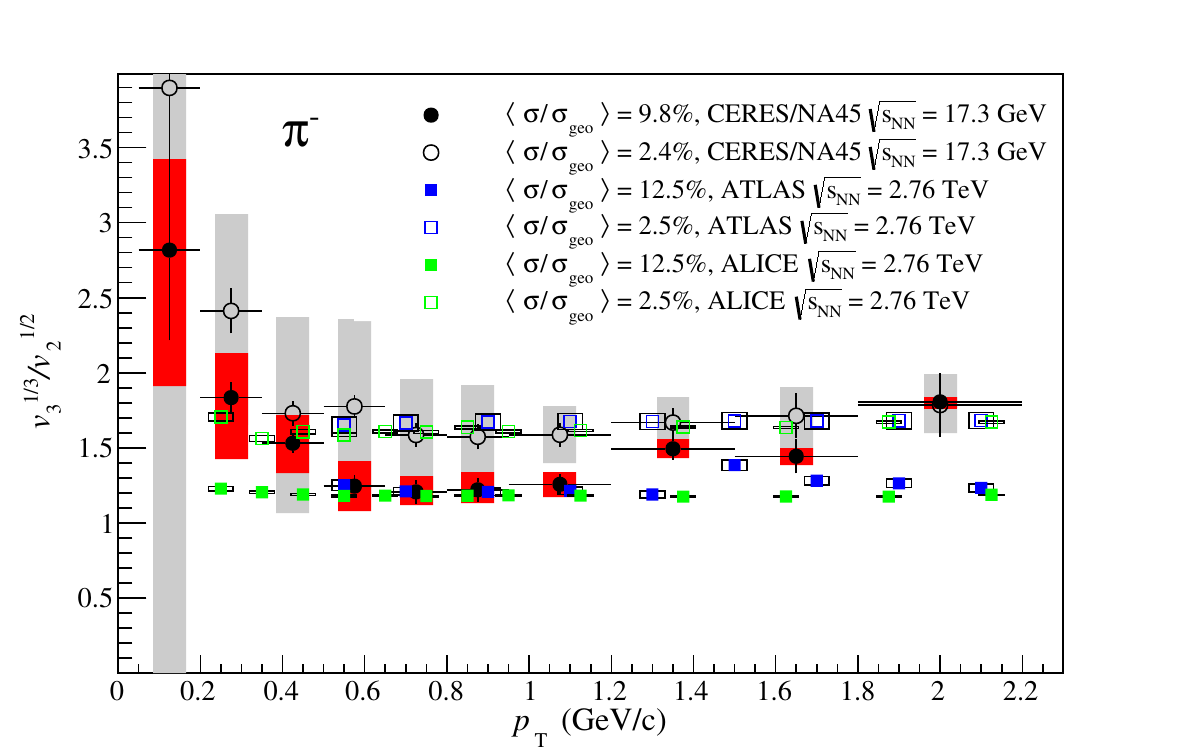}
\caption{The ratio $v^{1/3}_{3}/v^{1/2}_{2}$ for negative pions as a function of transverse momentum in PbAu collisions at $\sqrt{s_{\mathrm{NN}}} = $ 17.3~GeV in two centrality classes: `top-central' and `mid-central' events presented with open and closed circles respectively. The statistical uncertainties are represented as the vertical bars, while systematic ones are indicated by red and gray rectangles. The blue and green squares represent corresponding ATLAS~\cite{ATLAS:2012at} and ALICE~\cite{ALICE:2018rtz} results. Open boxes represent corresponding systematic uncertainties.
  \label{fig:4}}
  \end{figure}

Figure~\ref{fig:4} shows the ratio $v^{1/3}_{3}/v^{1/2}_{2}$ as a function of $p_{\mathrm{T}}$ for two centrality classes. The ratio is calculated for negative pions after suppression of the contribution of the HBT effect to the measured $v_{n}$ coefficients. The figure also shows the same ratio for PbPb collisions at $\sqrt{s_{\mathrm{NN}}}$= 2.76~TeV from the ATLAS~\cite{ATLAS:2012at} and ALICE~\cite{ALICE:2018rtz} experiment. The centrality classes are similar to those used by the CERES/NA45 experiment. The ATLAS and ALICE analyses have been performed on charged particles emitted within $|\eta| < 2.5$ and $|\eta| < 0.8$, respectively. There is a very good agreement between the ATLAS (blue squares) and ALICE (green squares) data. Despite the huge difference in incident energies between the top SPS energy and the LHC energy, for $p_{\mathrm{T}}$ above 0.5~GeV/c, the CERES/NA45 data are in a rather good agreement with the ATLAS and ALICE data, especially for the ‘top-central’ events. In the case of `mid-central' events for $p_{\mathrm{T}} >$~1.2~GeV/c, the CERES/NA45 results are somewhat above the ATLAS and ALICE data. Universally, there is a trend that the ratio for semicentral collisions is smaller. For the present data, there is a rising trend below 500~MeV/c, although not very significant within errors. The uncertainties are significantly larger for the current SPS data due to very small values of $v_{2,3}$ at the lower energy. In the low-$p_{\mathrm{T}}$ range 0.05 -- 0.5~GeV/c, where the HBT influence is largest, it may be that the HBT effect on the measured $v_{3}$ harmonic is not completely suppressed, thus producing larger values of the $v^{1/3}_{3}/v^{1/2}_{2}$ ratio. A $p_{\mathrm{T}}$-independent $v^{1/3}_{3}/v^{1/2}_{2}$ ratio is a hydrodynamic prediction~\cite{Alver:2010dn,Borghini:2005kd}. The CERES/NA45 results imply that the system formed in PbAu collisions at the top SPS energy exhibits hydrodynamic behavior as already suggested in~\cite{CERES:2012hbp}.

\begin{figure}[h!]
  \centerline{\includegraphics[width=9.2cm]{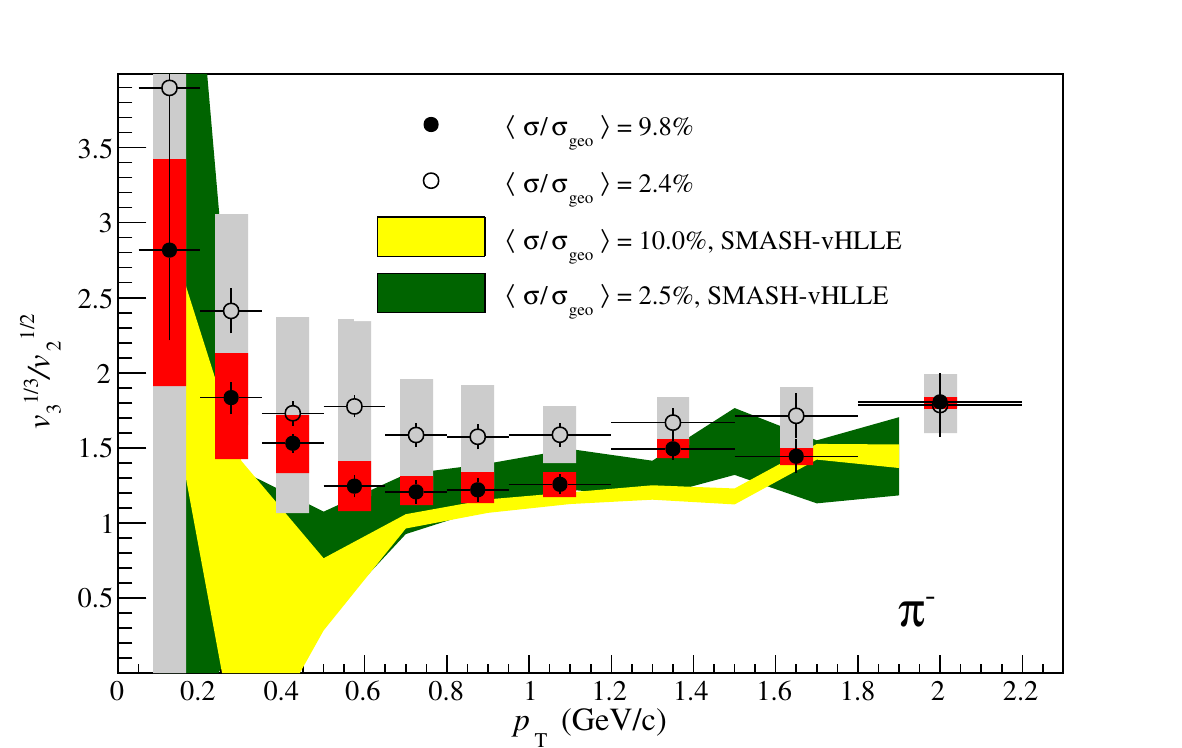}}
  \caption{The ratio $v^{1/3}_{3}/v^{1/2}_{2}$ for negative pions as a function of transverse momentum in PbAu collisions at $\sqrt{s_{\mathrm{NN}}} = $ 17.3~GeV in two centrality classes: `top-central' and `mid-central' events presented with open and closed circles respectively. The statistical uncertainties are represented with the vertical bars, while systematic ones are indicated by red and gray rectangles. Statistical uncertainties are represented with the vertical bars. The statistical uncertainties of the model predictions are shown as a yellow and green band.
\label{fig:5}}
\end{figure} 

Figure~\ref{fig:5} shows a comparison of the CERES $v^{1/3}_{3}/v^{1/2}_{2}$ data with the corresponding ratio calculated by the SMASH-vHLLE model. As expected, for $p_{\mathrm{T}} >$ 0.7 GeV/c, where the statistical uncertainties of the model are small enough, the model gives a rather flat ratio for both centrality classes. For $p_{\mathrm{T}} <$ 0.5 GeV/c, the statistical uncertainties of the model are very large. The SMASH-vHLLE model prediction tends to stay slightly below the experimental data. This is due to the fact that the model overestimates the experimentally measured $v_{2}$ somewhat for $p_{\mathrm{T}} >$ 0.7 GeV/c, while it reproduces $v_{3}$ harmonics, as shown in Figs.~\ref{fig:2} and \ref{fig:3}.

This possibility was already discussed in our first publication of elliptic flow coefficients for pions where data were compared to ideal hydrodynamics calculations~\cite{CERESNA45:2003pat} as well as in~\cite{CERES:2012hbp}. Then we found that data could be well decsribed using ideal hydrodynamics, and a freeze-out temperature of 160~MeV, while for 120~MeV the flow was overpredicted for semicentral data. At the time we concluded that a supression mechanism should be at work not present in ideal hydrodynamic calculations. This is confirmed by the rather good description of the data presented in the current publication with a viscous hydrodynamics calculation. Albeit the comparison presented here is to results of a more recent hydrodynamics code, the supression is in magnitude entirely consistent with the value of $\eta/s = $ 0.15 used here.

These predictions of the hybrid model at SPS tend to be a little below data. In particular, the choice of initial state and $\eta/s$ value for hydrodynamic evolution is very important. This measurement could help to better constrain the initial conditions and specific viscosity.

\section{Summary}
\label{summary}

The ratio $v^{1/3}_{3}/v^{1/2}_{2}$ of negative pions has been measured as a function of $p_{\mathrm{T}}$ using the data collected by the CERES/NA45 experiment for PbAu collisions at $\sqrt{s_{\mathrm{NN}}} = $ 17.3~GeV. This ratio, measured in two centrality classes, shows rather similar as the LHC results from the ATLAS and ALICE experiments. The SMASH-vHLLE hybrid model calculations reasonably well represent the data. Only for $p_{\mathrm{T}}$ above about 0.6 GeV/c, the SMASH-vHLLE hybrid model overestimates the experimentally measured $v_{2}$. The results give a new interpretation of the SPS data at its top energy which indicates hydrodynamic behavior of the system formed in PbAu collisions. These results shed some light on the dynamics of the system created in heavy-ion collisions at top SPS energy. The moderate deviations between data and the current hybrid hydrodynamics calculation could be used in future investigations to fine-tune the initial condition and transport parameters sofar rather little explained at SPS energy.

\begin{acknowledgements}
The CERES/NA45 collaboration acknowledges the good performance of the CERN PS and SPS accelerators as well as the support from the EST division. We are grateful for excellent support by the CERN IT division for the central data recording and data proccesing. This work was supported by GSI, Darmstadt, the German BMBF, the German VH-VI 146, the US DoE, the Israeli Science Foundation, and the MINERVA Foundation. We acknowledge the support by the Ministry of Education, Science and Technological Development of the Republic of Serbia throughout the theme 0102202. Z.P. acknowledges support by the State of Hesse within the Research Cluster ELEMENTS (Project ID 500/10.006). Computational resources for theoretical modelling have been provided by the GreenCube at GSI. We thank Iurii Karpenko for enlightening discussions with him, and appreciate his help concerning the SMASH-vHLLE hybrid model.
\end{acknowledgements}



\end{document}